\documentclass[dvips]{article}
\usepackage{supertabular,lscape,epsfig}
\usepackage{amssymb}
\usepackage{amsmath}

\DeclareSymbolFont{ppa}{OT1}{ppl}{m}{it}
\DeclareMathSymbol{\vv}{\mathalpha}{ppa}{'166}

\begin{document}

\newcommand{\dd}{\,{\rm d}}
\newcommand{\ie}{{\it i.e.},\,}
\newcommand{\etal}{{\it et al.\ }}
\newcommand{\eg}{{\it e.g.},\,}
\newcommand{\cf}{{\it cf.\ }}
\newcommand{\vs}{{\it vs.\ }}
\newcommand{\zdot}{\makebox[0pt][l]{.}}
\newcommand{\up}[1]{\ifmmode^{\rm #1}\else$^{\rm #1}$\fi}
\newcommand{\dn}[1]{\ifmmode_{\rm #1}\else$_{\rm #1}$\fi}
\newcommand{\upd}{\up{d}}
\newcommand{\uph}{\up{h}}
\newcommand{\upm}{\up{m}}
\newcommand{\ups}{\up{s}}
\newcommand{\arcd}{\ifmmode^{\circ}\else$^{\circ}$\fi}
\newcommand{\arcm}{\ifmmode{'}\else$'$\fi}
\newcommand{\arcs}{\ifmmode{''}\else$''$\fi}
\newcommand{\MS}{{\rm M}\ifmmode_{\odot}\else$_{\odot}$\fi}
\newcommand{\RS}{{\rm R}\ifmmode_{\odot}\else$_{\odot}$\fi}
\newcommand{\LS}{{\rm L}\ifmmode_{\odot}\else$_{\odot}$\fi}

\newcommand{\Abstract}[2]{{\footnotesize\begin{center}ABSTRACT\end{center}
\vspace{1mm}\par#1\par
\noindent
{~}{\it #2}}}

\newcommand{\TabCap}[2]{\begin{center}\parbox[t]{#1}{\begin{center}
  \small {\spaceskip 2pt plus 1pt minus 1pt T a b l e}
  \refstepcounter{table}\thetable \\[2mm]
  \footnotesize #2 \end{center}}\end{center}}

\newcommand{\TableSep}[2]{\begin{table}[p]\vspace{#1}
\TabCap{#2}\end{table}}

\newcommand{\FigCap}[1]{\footnotesize\par\noindent Fig.\  %
  \refstepcounter{figure}\thefigure. #1\par}

\newcommand{\TableFont}{\footnotesize}
\newcommand{\TableFontIt}{\ttit}
\newcommand{\SetTableFont}[1]{\renewcommand{\TableFont}{#1}}

\newcommand{\MakeTable}[4]{\begin{table}[htb]\TabCap{#2}{#3}
  \begin{center} \TableFont \begin{tabular}{#1} #4 
  \end{tabular}\end{center}\end{table}}

\newcommand{\MakeTableSep}[4]{\begin{table}[p]\TabCap{#2}{#3}
  \begin{center} \TableFont \begin{tabular}{#1} #4 
  \end{tabular}\end{center}\end{table}}

\newenvironment{references}%
{
\footnotesize \frenchspacing
\renewcommand{\thesection}{}
\renewcommand{\in}{{\rm in }}
\renewcommand{\AA}{Astron.\ Astrophys.}
\newcommand{\AAS}{Astron.~Astrophys.~Suppl.~Ser.}
\newcommand{\ApJ}{Astrophys.\ J.}
\newcommand{\ApJS}{Astrophys.\ J.~Suppl.~Ser.}
\newcommand{\ApJL}{Astrophys.\ J.~Letters}
\newcommand{\AJ}{Astron.\ J.}
\newcommand{\IBVS}{IBVS}
\newcommand{\PASP}{P.A.S.P.}
\newcommand{\Acta}{Acta Astron.}
\newcommand{\MNRAS}{MNRAS}
\renewcommand{\and}{{\rm and }}
\section{{\rm REFERENCES}}
\sloppy \hyphenpenalty10000
\begin{list}{}{\leftmargin1cm\listparindent-1cm
\itemindent\listparindent\parsep0pt\itemsep0pt}}%
{\end{list}\vspace{2mm}}

\def\TYLDA{~}
\newlength{\DW}
\settowidth{\DW}{0}
\newcommand{\dw}{\hspace{\DW}}

\newcommand{\refitem}[5]{\item[]{#1} #2%
\def\REFARG{#3}\ifx\REFARG\TYLDA\else, {\it#3}\fi
\def\REFARG{#4}\ifx\REFARG\TYLDA\else, {\bf#4}\fi
\def\REFARG{#5}\ifx\REFARG\TYLDA\else, {#5}\fi.}

\newcommand{\Section}[1]{\section{#1}}
\newcommand{\Subsection}[1]{\subsection{#1}}
\newcommand{\Acknow}[1]{\par\vspace{5mm}{\bf Acknowledgements.} #1}
\pagestyle{myheadings}

\newfont{\bb}{ptmbi8t at 12pt}
\newcommand{\xrule}{\rule{0pt}{2.5ex}}
\newcommand{\xxrule}{\rule[-1.8ex]{0pt}{4.5ex}}
\def\thefootnote{\fnsymbol{footnote}}
\begin{center}
{\Large\bf The Optical Gravitational Lensing Experiment.\\
\vskip3pt
{BVI} Maps of Dense Stellar Regions.\\
\vskip6pt
II. The Large Magellanic Cloud\footnote{Based on  observations obtained
with the 1.3~m Warsaw telescope at the Las Campanas  Observatory of the
Carnegie Institution of Washington.}}

\vskip1cm
{\bf A.~~U~d~a~l~s~k~i$^1$,~~M.~~S~z~y~m~a~{\'n}~s~k~i$^1$,~~
M.~~K~u~b~i~a~k$^1$,\\ 
G.~~P~i~e~t~r~z~y~\'n~s~k~i$^{1,2}$,~~ I.~~S~o~s~z~y~\'n~s~k~i,~~
P.~~W~o~\'z~n~i~a~k$^3$,
~~ and~~K.~~\.Z~e~b~r~u~\'n$^1$}
\vskip3mm
{$^1$Warsaw University Observatory, Al.~Ujazdowskie~4, 00-478~Warszawa,
Poland\\
e-mail: (udalski,msz,mk,pietrzyn,soszynsk,zebrun)@astrouw.edu.pl\\
$^2$ Universidad de Concepci{\'o}n, Departamento de Fisica,
Casilla 160--C, Concepci{\'o}n, Chile\\
$^3$ Princeton University Observatory, Princeton, NJ 08544-1001, USA\\
e-mail: wozniak@astro.princeton.edu}
\end{center}

\Abstract{We present the {\it BVI} photometric maps of the Large Magellanic 
Cloud. They contain {\it BVI} photometry and astrometry of more than 7 
million stars from the central parts of the LMC. The data were collected 
during the second phase of the OGLE microlensing project. We discuss the 
accuracy of the data and present color-magnitude diagrams of all 26 fields 
observed by OGLE in the LMC. 

The {\it BVI} maps of the LMC are accessible electronically for the 
astronomical community from the OGLE Internet archive.}

\Section{Introduction}
The Large Magellanic Cloud (LMC) is one of the most important astrophysical 
objects. This nearby galaxy hosts large variety of very important stellar 
populations and is an ideal laboratory for testing our understanding of 
stellar structure and evolution and for calibrating the most important 
standard candles. The LMC is also very important object for extragalactic 
astrophysics -- the extragalactic distance scale is calibrated by the distance 
to the LMC. Due to relatively small distance from the Galaxy, stars and other 
objects from the LMC can easily be resolved and measured. Approximately the 
same distance to all objects from that galaxy makes it especially attractive 
for determination and comparison of basic parameters of different groups of 
stars. 

Unfortunately, the LMC was relatively poorly observed, in particular with 
precise CCD techniques. Photometry of only selected fields in lines of sight 
usually toward star clusters in the halo of this galaxy can be found in 
literature. The situation considerably changed when the large microlensing 
surveys began regular monitoring of the Magellanic Clouds for microlensing 
events. The LMC is one of the most promising targets of the microlensing hunt, 
and therefore it was selected as a target of MACHO (Alcock \etal 2000) and 
EROS (Lasserre \etal 2000) microlensing projects. Large areas in the LMC were 
also observed by Zaritsky, Harris and Thompson (1997) as a part of a 
photometric survey of the Magellanic Clouds. 

The LMC was included as one of the targets of the Optical Gravitational 
Lensing Experiment (OGLE) at the beginning of the second phase of the project 
(Udalski, Kubiak and Szyma{\'n}ski 1997). Since January 1997 the large part of 
the central regions of the galaxy has been regularly observed practically on every 
clear night. As a result huge databases containing hundreds of photometric 
measurements of millions of stars were created. 

The OGLE photometric data are particularly attractive for many projects 
requiring precise photometry. Among the other datasets collected during the 
microlensing searches, only the OGLE photometry was obtained with the standard 
{\it BVI} filters what makes it very well suited for projects not related to 
microlensing. Also, very good astronomical site where the OGLE project is 
conducted -- the Las Campanas Observatory in Chile, made it possible to 
achieve the best resolution what is crucial in the very dense stellar fields 
of the LMC bar. 

In this paper, which is the second of the series, we present "The OGLE {\it 
BVI} Maps of the LMC." The maps contain the mean {\it BVI} photometry and 
astrometry of more than 7 million stars from the field of about 5.7~square 
degrees in the central part of the LMC. With the maps of the Small Magellanic 
Cloud presented in the first paper of this series (Udalski \etal 1998b) they 
constitute a unique photometric database of objects from the Magellanic 
Clouds. 

Because of potentially great impact on many astrophysical fields, the OGLE 
policy is to make the photometric data available to the astronomical 
community. Similarly to the SMC maps, the maps of the LMC are also available 
from the OGLE Internet archive. Details are provided in the last Section of 
this paper. 

\Section{Observations}
All observations presented in this paper were collected during the second 
phase of the OGLE microlensing search with the 1.3-m Warsaw telescope at the 
Las Campanas Observatory, Chile, which is operated by the Carnegie Institution 
of Washington. The telescope was equipped with the "first generation" camera 
with  a SITe ${2048\times2048}$ CCD detector working in drift-scan mode. The 
pixel size was 24~$\mu$m giving the 0.417 arcsec/pixel scale. Observations 
were performed in the "slow" reading mode of the CCD detector with the gain 
3.8~e$^-$/ADU and readout noise of about 5.4~e$^-$. Details of the 
instrumentation setup can be found in Udalski, Kubiak and Szyma{\'n}ski 
(1997). 
\MakeTable{lcc}{12.5cm}{Equatorial coordinates of the OGLE-II LMC fields}
{
\hline
\noalign{\vskip3pt}
\multicolumn{1}{c}{Field} & RA (J2000) & DEC (J2000)\\
\hline
\noalign{\vskip3pt}
LMC$\_$SC1  &  5\uph33\upm49\ups & $-70\arcd06\arcm10\arcs$ \\
LMC$\_$SC2  &  5\uph31\upm17\ups & $-69\arcd51\arcm55\arcs$ \\
LMC$\_$SC3  &  5\uph28\upm48\ups & $-69\arcd48\arcm05\arcs$ \\
LMC$\_$SC4  &  5\uph26\upm18\ups & $-69\arcd48\arcm05\arcs$ \\
LMC$\_$SC5  &  5\uph23\upm48\ups & $-69\arcd41\arcm05\arcs$ \\
LMC$\_$SC6  &  5\uph21\upm18\ups & $-69\arcd37\arcm10\arcs$ \\
LMC$\_$SC7  &  5\uph18\upm48\ups & $-69\arcd24\arcm10\arcs$ \\
LMC$\_$SC8  &  5\uph16\upm18\ups & $-69\arcd19\arcm15\arcs$ \\
LMC$\_$SC9  &  5\uph13\upm48\ups & $-69\arcd14\arcm05\arcs$ \\
LMC$\_$SC10 &  5\uph11\upm16\ups & $-69\arcd09\arcm15\arcs$ \\
LMC$\_$SC11 &  5\uph08\upm41\ups & $-69\arcd10\arcm05\arcs$ \\
LMC$\_$SC12 &  5\uph06\upm16\ups & $-69\arcd38\arcm20\arcs$ \\
LMC$\_$SC13 &  5\uph06\upm14\ups & $-68\arcd43\arcm30\arcs$ \\
LMC$\_$SC14 &  5\uph03\upm49\ups & $-69\arcd04\arcm45\arcs$ \\
LMC$\_$SC15 &  5\uph01\upm17\ups & $-69\arcd04\arcm45\arcs$ \\
LMC$\_$SC16 &  5\uph36\upm18\ups & $-70\arcd09\arcm40\arcs$ \\
LMC$\_$SC17 &  5\uph38\upm48\ups & $-70\arcd16\arcm45\arcs$ \\
LMC$\_$SC18 &  5\uph41\upm18\ups & $-70\arcd24\arcm50\arcs$ \\
LMC$\_$SC19 &  5\uph43\upm48\ups & $-70\arcd34\arcm45\arcs$ \\
LMC$\_$SC20 &  5\uph46\upm18\ups & $-70\arcd44\arcm50\arcs$ \\
LMC$\_$SC21 &  5\uph21\upm14\ups & $-70\arcd33\arcm20\arcs$ \\
LMC$\_$SC22 &  5\uph02\upm26\ups & $-67\arcd09\arcm35\arcs$ \\
LMC$\_$SC23 &  5\uph04\upm45\ups & $-67\arcd09\arcm40\arcs$ \\
LMC$\_$SC24 &  5\uph07\upm05\ups & $-67\arcd09\arcm35\arcs$ \\
LMC$\_$SC25 &  5\uph09\upm24\ups & $-67\arcd09\arcm30\arcs$ \\
LMC$\_$SC26 &  5\uph11\upm43\ups & $-67\arcd09\arcm40\arcs$ \\
\hline}

Observed fields covered practically the entire bar of the LMC. More than 4.5 
square degrees (21 driftscan fields) were monitored regularly (practically on 
every clear night except for the May--August period each year) from January 
1997 through May 2000. Five additional fields in the North--West part of the 
LMC were monitored on 13 nights between November 1998 and January 1999. 
Table~1 lists the equatorial coordinates of the centers of each field 
($14.2\times57$ arcmin)~with their acronyms. Positions of the fields were 
chosen in that way that adjacent fields overlap by about one arcmin for 
testing purposes. Fig.~1 presents the picture of the LMC from the Digitized 
Sky Survey with contours of the OGLE-II fields. 

Observations were obtained with the standard {\it BVI} filters closely 
reproducing the standard system (Section~3). Due to the microlensing search 
observing strategy, the vast majority of observations were obtained through 
the {\it I}-band filter (250--500 per field) while much smaller number of 
frames in the {\it BV}-bands were collected (25--40 and 40--80 for the {\it B} 
and {\it V} filters, respectively). For the additional North--West fields only 
6 and 7 observations in the {\it V} and {\it I}-bands were collected. The 
effective exposure time was 125, 174 and 237 seconds, for the {\it I, V} and 
{\it B}-band, respectively. 

Altogether more than 8800 images (about 300~GB of raw data) of the OGLE-II LMC 
fields were collected during the presented period of observations. Because of 
high stellar density of the bar fields, observations were only conducted 
during the nights with good seeing conditions. The median seeing of the entire 
data set is about 1.3~arcsec. Observations of the LMC were usually stopped 
when the seeing exceeded 1.6--1.8~arcsec. 

\Section{Data Reduction and Calibration}
All collected frames were reduced using the standard OGLE data pipeline in the 
identical manner as the SMC data (Udalski \etal 1998b). The data pipeline is 
described in detail in Udalski \etal (1998b). In short, after de-biasing and 
flat-fielding, photometry of objects is derived using the {\sc DoPhot} 
photometry program (Schechter, Saha and Mateo 1993) running in the fixed position 
mode on sixty four ${512\times 512}$~pixel subframes. The full 
${2048\times8192}$~pixel image being reduced is first matched with the so 
called "template" image, \ie the image with very good angular resolution 
(obtained at very good seeing conditions) and then divided into subframes. 
Photometry of each subframe is tied to the photometry of the template subframe 
by adding the mean shift derived usually from several hundreds brighter stars. 
Thus, photometry of the template image defines the instrumental system. Objects 
detected in the template images for the {\it B} and {\it V}-bands are first 
matched with the {\it I}-band template image objects of a given field so the 
star numbering is the same in all bands making the data handling much easier. 
Due to small shifts of the {\it BV} template images in respect to the 
{\it I}-band image not all stars detected in the {\it I}-band image have full 
{\it BVI} photometry. However, they usually have full photometry in the 
adjacent overlapping field. 

To determine transformations of the instrumental photometry to the standard 
system, several Landolt (1992) standard fields were observed on about 250 
photometric nights during the OGLE-II observations. Based on thousands of 
observations of standard stars in Landolt (1992) fields located all over the 
sky and covering large range of colors, the following average transformations 
were derived: 
\begin{eqnarray}
B  &=&b-0.034\times(B-V) +{\rm const}_B\nonumber\\
V  &=&\vv-0.002\times(V-I) +{\rm const}_V\nonumber\\
I  &=&i+0.029\times(V-I) +{\rm const}_I\\
B-V&=& 0.963\times(b-\vv)+{\rm const}_{B-V}\nonumber\\
V-I&=& 0.969\times(\vv-i)+{\rm const}_{V-I}\nonumber
\end{eqnarray}

The typical residuals of calculated minus observed magnitudes of standard 
stars did not exceed 0.03~mag. The transformations indicate that the OGLE-II 
filter system closely resembles the standard one, \ie the color coefficients 
are close to zero or one. Observations of standard stars also indicate that 
the instrumental system was extremely stable during the period of observations 
and the standard system magnitudes could be derived with good accuracy 
(${0.02-0.04}$~mag) even during the photometric nights when no standard 
stars were observed. 

Transformation of the instrumental magnitudes to the standard system consisted 
of the following steps. First, the aperture corrections were determined for 
each of the 64 subframes. They were determined from aperture photometry of 
typically 20--100 stars per subframe measured in the artificial images with 
faint stars subtracted. Then the total correction was determined consisting of 
the aperture correction, zero point of transformation, extinction correction 
and normalization to 1~sec exposure time. The total corrections were 
determined independently for about 10--25, 10--20 and 20--35 photometric 
nights for the {\it BV} and {\it I}-band, respectively. Typical standard 
deviation of the total correction in each of the 64 subframes was of about 
0.015--0.020~mag. The averaged values of the total correction were 
subsequently used for the construction of photometric databases (Udalski \etal 
1998b, Szyma{\'n}ski and Udalski 1993) for the {\it B, V} and {\it I}-band. 
The databases contain entire photometry of all objects in a given OGLE field 
in the system very close to the standard one -- only the color term (Eq.~1) is 
not included. 

Equatorial coordinates of objects detected in the OGLE fields were determined 
in the identical manner as described in Udalski \etal (1998b). Objects in the 
OGLE-II frames were cross-identified with the objects detected in the 
Digitized Sky Survey images, and the transformation between the OGLE pixel 
grid and (RA,DEC) coordinates in the DSS coordinate system was determined. 
About 1700--7400 stars were used for transformation depending on the stellar 
density in the field. Internal accuracy of the determined equatorial 
coordinates, as measured in the overlapping regions of neighboring fields is 
about 0.15--0.20~arcsec. However, the systematic error of the DSS coordinate 
system may reach 0.7~arcsec. 
\vspace*{10pt}
\Section{{\bb BVI} Maps of the LMC}
\vspace*{5pt}
The {\it BVI} maps of the LMC were constructed using the photometric databases 
of each field. First, the mean magnitudes of each object were 
calculated with $5\sigma$ rejection alghoritm. Then, we corrected the 
magnitudes for a small systematic error, caused by non-perfect flat-fielding 
at the edges of the field. This effect was first noticed by Dr.\ D.S.\ Graff 
(Ohio State University), and it was precisely mapped based on observations of 
hundreds of standard stars. Finally, the color corrections (Eq.~1) were 
derived and added. Only objects with more than 5, 10 and 40 good observations 
(see Udalski \etal 1998b) in the {\it BVI}-bands, respectively, were included 
in the final maps of the LMC (three in the {\it V} and {\it I}-band in the 
North--West LMC fields). Table~2 lists the total number of objects in the 
OGLE-II maps of the LMC fields. 
\MakeTable{lcclc}{12.5cm}{Number of objects in the OGLE-II LMC maps}
{
\cline{1-2}\cline{4-5}
\noalign{\vskip3pt}
\multicolumn{1}{c}{Field} & $N_{\rm objects}$ &~~~~~~&
\multicolumn{1}{c}{Field} & $N_{\rm objects}$ \\
\noalign{\vskip3pt}
\cline{1-2}\cline{4-5}
\noalign{\vskip3pt}
LMC$\_$SC1  &  348876 &~~~~~~ &LMC$\_$SC14 &  263522\\  
LMC$\_$SC2  &  421346 &~~~~~~ &LMC$\_$SC15 &  223214\\ 
LMC$\_$SC3  &  448729 &~~~~~~ &LMC$\_$SC16 &  268601\\ 
LMC$\_$SC4  &  485406 &~~~~~~ &LMC$\_$SC17 &  238924\\ 
LMC$\_$SC5  &  460020 &~~~~~~ &LMC$\_$SC18 &  211320\\ 
LMC$\_$SC6  &  474475 &~~~~~~ &LMC$\_$SC19 &  194556\\ 
LMC$\_$SC7  &  477838 &~~~~~~ &LMC$\_$SC20 &  209089\\ 
LMC$\_$SC8  &  368108 &~~~~~~ &LMC$\_$SC21 &  198156\\ 
LMC$\_$SC9  &  397024 &~~~~~~ &LMC$\_$SC22 &  ~~68061\\
LMC$\_$SC10 &  292552 &~~~~~~ &LMC$\_$SC23 &  ~~71475\\
LMC$\_$SC11 &  352465 &~~~~~~ &LMC$\_$SC24 &  ~~71012\\
LMC$\_$SC12 &  215090 &~~~~~~ &LMC$\_$SC25 &  ~~60276\\
LMC$\_$SC13 &  272614 &~~~~~~ &LMC$\_$SC26 &  ~~54083\\
\cline{1-2}\cline{4-5}
\noalign{\vskip3pt}     
&&& Total: & 7146832      
}                 

Table~3 presents the sample data from the map of the LMC$\_$SC3 field. In the 
consecutive columns the following data are provided: star ID number, 
equatorial coordinates, ($X,Y$) coordinates in the {\it I}-band template image, 
photometry: {\it V}, ${(B-V)}$, ${(V-I)}$, {\it B}, {\it I}, number of 
observations, number of rejected observations and standard deviation for the 
{\it BVI}-bands, respectively. In the electronic version we additionally 
provide the FITS template images for easy object identification. In the case 
of the North--West LMC fields only {\it V} and {\it I} images were collected. 
Therefore, the maps of these fields contain dummy values in the columns 
related to {\it B}-band photometry. Maps of all LMC fields are available 
electronically from the OGLE Internet archive (see Section~7). 

\Section{Data Tests} 
\Subsection{Photometry}
Quality of the OGLE-II photometry can be assessed from comparison of 
magnitudes of stars located in the overlaps between neighboring fields. 
Because each of the fields was calibrated independently such a comparison 
provides information on accuracy of calibration and the typical accuracy of 
photometry. Figs.~2, 3 and 4 present differences of magnitudes for stars with 
magnitudes brighter than ${B=20}$~mag, ${V=20}$~mag and ${I=19}$~mag, 
plotted as a function of line number for three fields located in the central 
part of the LMC and at the West and East edges of the bar. For the remaining 
fields the plots look very similar. As can be seen the average difference of 
magnitudes is always below 0.01~mag indicating good consistency of the 
calibration procedure. The typical sigma of the Gaussian fitted to the 
histogram of differences of magnitudes is about 0.040, 0.035 and 0.025~mag for 
the {\it BVI}-bands, respectively. 
\begin{landscape}
\MakeTable{c@{\hspace{5pt}}c@{\hspace{5pt}}c@{\hspace{7pt}}r
@{\hspace{7pt}}r@{\hspace{7pt}}c@{\hspace{5pt}}c@{\hspace{5pt}}c
@{\hspace{5pt}}c@{\hspace{5pt}}c@{\hspace{5pt}}c@{\hspace{5pt}}c
@{\hspace{5pt}}c@{\hspace{5pt}}c@{\hspace{5pt}}c@{\hspace{5pt}}c
@{\hspace{5pt}}r@{\hspace{5pt}}c@{\hspace{5pt}}c}{12.5cm}{Sample 
of data from the {\it BVI} map of the field LMC$\_$SC3}
{
\hline
\noalign{\vskip5pt}
Star & RA & DEC & \multicolumn{1}{c}{$X$} & \multicolumn{1}{c}{$Y$} & $V$ & $B-V$ & $V-I$ & $B$ & 
$I$ & $N_{\rm ok}^B$ & $N_{\rm bad}^B$ & $\sigma_B$ & $N_{\rm ok}^V$ & $N_{\rm bad}^V$ & 
$\sigma_V$ & $N_{\rm ok}^I$ & $N_{\rm bad}^I$ & $\sigma_I$\\
no & (J2000) & (J2000) &&&&&&&&&&&&&&&&\\
\noalign{\vskip5pt}
\hline
\noalign{\vskip5pt}
1  & 5\uph27\upm46\zdot\ups52 & $-70\arcd15\arcm47\zdot\arcs6$ & 275.31 &  77.35 & 13.147 & 0.557 & 0.689 & 13.709 & 12.459 & 21 & 0 & 0.012 & 52 & 1 & 0.016 & 382 & 1 & 0.031\\
2  & 5\uph28\upm03\zdot\ups66 & $-70\arcd15\arcm46\zdot\arcs7$ & 485.35 &  80.59 & 16.241 & 2.948 & 2.596 & 19.208 & 13.643 & 12 & 0 & 0.522 & 56 & 0 & 0.307 & 382 & 0 & 0.138\\
3  & 5\uph27\upm45\zdot\ups67 & $-70\arcd16\arcm01\zdot\arcs5$ & 265.17 &  43.68 & 16.831 & 1.795 & 2.888 & 18.641 & 13.940 & 23 & 0 & 0.184 & 58 & 0 & 0.171 & 326 & 0 & 0.071\\
4  & 5\uph28\upm03\zdot\ups60 & $-70\arcd15\arcm25\zdot\arcs4$ & 484.35 & 131.92 & 16.762 & 1.713 & 2.338 & 18.488 & 14.423 & 23 & 0 & 0.084 & 56 & 0 & 0.063 & 406 & 2 & 0.025\\
5  & 5\uph27\upm50\zdot\ups65 & $-70\arcd15\arcm08\zdot\arcs9$ & 325.49 & 171.01 & 17.074 & 1.737 & 2.765 & 18.824 & 14.307 & 23 & 0 & 0.176 & 57 & 0 & 0.161 & 416 & 0 & 0.062\\
6  & 5\uph27\upm29\zdot\ups70 & $-70\arcd14\arcm49\zdot\arcs7$ &  68.34 & 215.98 & 16.934 & 4.410 & 2.856 & 21.368 & 14.076 & 11 & 0 & 0.314 & 50 & 0 & 0.215 & 394 & 0 & 0.170\\
7  & 5\uph27\upm24\zdot\ups73 & $-70\arcd14\arcm29\zdot\arcs1$ &   7.13 & 265.23 & 18.086 &    -- & 3.786 &  --    & 14.296 &  0 & 0 & --    & 29 & 0 & 0.265 & 318 & 0 & 0.140\\
8  & 5\uph27\upm38\zdot\ups43 & $-70\arcd13\arcm28\zdot\arcs2$ & 174.30 & 413.48 & 14.972 & 0.677 & 0.769 & 15.653 & 14.202 & 18 & 1 & 0.020 & 58 & 0 & 0.012 & 416 & 0 & 0.013\\
9  & 5\uph27\upm56\zdot\ups02 & $-70\arcd13\arcm20\zdot\arcs0$ & 390.14 & 434.47 & 16.565 & 1.847 & 2.326 & 18.425 & 14.236 & 23 & 0 & 0.054 & 58 & 0 & 0.034 & 419 & 0 & 0.018\\
10 & 5\uph27\upm35\zdot\ups32 & $-70\arcd13\arcm17\zdot\arcs6$ & 136.03 & 439.01 & 17.185 & 1.894 & 2.917 & 19.093 & 14.264 & 23 & 0 & 0.197 & 58 & 0 & 0.234 & 402 & 0 & 0.121\\
11 & 5\uph28\upm03\zdot\ups09 & $-70\arcd13\arcm08\zdot\arcs4$ & 476.77 & 463.04 & 15.957 & 1.745 & 2.331 & 17.715 & 13.624 & 23 & 0 & 0.080 & 58 & 0 & 0.050 & 402 & 0 & 0.026\\
12 & 5\uph27\upm51\zdot\ups54 & $-70\arcd13\arcm00\zdot\arcs1$ & 334.94 & 482.34 & 17.646 & 1.901 & 3.470 & 19.562 & 14.173 & 23 & 0 & 0.134 & 57 & 0 & 0.147 & 408 & 0 & 0.075\\
14 & 5\uph27\upm39\zdot\ups92 & $-70\arcd16\arcm10\zdot\arcs2$ & 194.78 &  22.33 & 16.419 & 1.705 & 1.749 & 18.134 & 14.668 &  6 & 1 & 0.012 & 57 & 0 & 0.023 & 263 & 0 & 0.018\\
15 & 5\uph27\upm43\zdot\ups91 & $-70\arcd16\arcm05\zdot\arcs6$ & 243.67 &  33.66 & 16.683 & 1.737 & 1.742 & 18.432 & 14.940 & 23 & 0 & 0.037 & 57 & 0 & 0.031 & 292 & 0 & 0.027\\
16 & 5\uph27\upm40\zdot\ups01 & $-70\arcd15\arcm53\zdot\arcs4$ & 195.66 &  62.91 & 15.218 & 0.662 & 0.771 & 15.884 & 14.446 & 18 & 0 & 0.013 & 54 & 0 & 0.016 & 389 & 2 & 0.017\\
17 & 5\uph27\upm34\zdot\ups48 & $-70\arcd15\arcm43\zdot\arcs2$ & 127.80 &  87.10 & 15.966 & 1.357 & 1.297 & 17.332 & 14.668 & 22 & 0 & 0.027 & 54 & 1 & 0.010 & 401 & 2 & 0.014\\
18 & 5\uph27\upm25\zdot\ups02 & $-70\arcd15\arcm33\zdot\arcs0$ &  11.66 & 111.04 & 16.766 &   --  & 1.641 &  --    & 15.123 &  0 & 0 & --    & 40 & 0 & 0.020 & 336 & 3 & 0.016\\
19 & 5\uph27\upm38\zdot\ups54 & $-70\arcd15\arcm23\zdot\arcs5$ & 177.21 & 134.87 & 16.792 & 1.626 & 1.578 & 18.429 & 15.212 & 23 & 0 & 0.045 & 56 & 0 & 0.011 & 414 & 2 & 0.012\\
20 & 5\uph27\upm51\zdot\ups46 & $-70\arcd15\arcm16\zdot\arcs8$ & 335.48 & 152.15 & 16.870 & 1.720 & 2.022 & 18.602 & 14.847 & 23 & 0 & 0.057 & 58 & 0 & 0.027 & 413 & 3 & 0.016\\
21 & 5\uph27\upm50\zdot\ups36 & $-70\arcd15\arcm12\zdot\arcs1$ & 322.02 & 163.25 & 16.762 & 1.751 & 1.800 & 18.524 & 14.960 & 23 & 0 & 0.063 & 58 & 0 & 0.025 & 406 & 0 & 0.018\\
22 & 5\uph27\upm24\zdot\ups41 & $-70\arcd15\arcm08\zdot\arcs8$ &   3.80 & 169.41 & 16.446 &   --  & 1.663 &  --    & 14.783 &  0 & 0 & --    & 33 & 1 & 0.030 & 274 & 0 & 0.021\\
\noalign{\vskip5pt}
\hline
}
\end{landscape}

The error of photometric measurements is a function of stellar magnitude. To 
illustrate that dependence Figs.~5 and 6 show the standard deviation of {\it 
B, V} and {\it I}-band magnitudes of objects in the central field LMC$\_$SC6 
and much less crowded LMC$\_$SC12. Only about 50~000 stars are plotted for 
clarity. 

Accuracy of the zero points of the OGLE photometry can be estimated from 
comparison of the OGLE data with other observations. Unfortunately, the number 
of good quality photometric data from the LMC bar is very limited. These 
regions were rarely observed in the past, in particular with precise CCD 
detectors. We have found in the literature only three cases we can compare our 
data with. 

Udalski \etal (1998a) presented comparison of the {\it I}-band light curves of 
Ce\-pheids located close to NGC~1850 with those obtained by Sebo and Wood 
(1995) indicating good agreement of the zero points of both data sets. Although 
the LMC data were only preliminarily calibrated at that time, the final 
calibration of that OGLE field is different only by a few thousandths of 
magnitude. Therefore the conclusion of Udalski \etal (1998a) fully holds. 

Walker (1993) presented {\it BV} photometry of stars in the field of NGC~1835. 
In addition to the photometry of variable stars he also published magnitudes 
of nine local standard stars. One of these stars (\#11) turned out to be a 
small amplitude long term variable as indicated by the OGLE data. One of the 
remaining eight stars (\#4) is brighter by about 0.25~mag than indicated by the 
OGLE photometry. Although the OGLE data do not show any larger light 
variations of this object, it is likely that so large discrepancy might be 
caused by some long term activity. Also OGLE data indicate that the star \#6 
is significantly fainter (more than 0.1~mag) than according to Walker. For the 
remaining six objects the mean difference of {\it V} magnitudes is equal to: 
${\Delta V=-0.026\pm0.025}$~mag indicating good agreement of both data sets. 
Walker's magnitudes are slightly brighter. Comparison of ${B-V}$ colors of 
Walker standards with the OGLE data indicates, however, that for stars redder 
than ${(B-V)\approx1.2}$~mag the OGLE colors are by more than 0.1~mag redder. 
For the four Walker standards outside that range the agreement is good -- the 
mean difference of $(B-V)$ is equal to ${\Delta(B-V)=-0.026\pm0.028}$~mag with 
Walker's colors being bluer. The discrepancy for redder stars is somewhat 
worrying. However, the Walker {\it B}-band filter rather poorly matched the 
standard {\it B}-band (color coefficient of transformation of about 1.2; see Walker 1992) and 
therefore larger systematic color differences for stars with extreme red 
colors are very likely. Apart from this problem the overall agreement between 
the Walker's and OGLE {\it BV} photometry calibrations is quite good. 

Finally, Clementini \etal (2000) presented photometry of two fields 
located close to the LMC bar. One of their fields, Field~A, overlaps in large 
part with the OGLE field LMC$\_$SC21. Unfortunately, a couple of local 
standard stars for which {\it BV} photometry is presented by Clementini \etal 
(2000) are located outside the OGLE field. Therefore the only comparison with 
the OGLE photometry which can be made is based on the average magnitudes of 
groups of stars, namely RR~Lyr and red clump stars. 

We extracted RR~Lyr and red clump stars located in the LMC$\_$SC21 field and 
limited these samples to objects with declination within the Clementini \etal 
(2000) Field~A limits. One should be aware that the two compared regions are 
not exactly the same, but because the interstellar extinction is quite uniform 
in this part of the LMC it is very unlikely to introduce any significant error 
by limiting the OGLE field in declination only. On the other hand much larger 
sample of stars can be averaged (in particular RR~Lyr stars) making the mean 
magnitudes much more reliable. 

For RR~Lyr stars (71 objects) the intensity mean magnitudes were first 
derived. Then the average magnitudes of both RR~Lyr and red clump stars were 
calculated. Comparison with Clementini \etal (2000) values yields: RR~Lyr -- 
${\langle V\rangle=19.35}$~mag \vs ${\langle V\rangle=19.37}$~mag and 
${\langle B-V\rangle=0.41}$~mag \vs ${\langle B-V\rangle=0.37}$ mag, for 
Clementini \etal (2000) and OGLE, respectively. For red clump stars -- 
${\langle V\rangle=19.22}$~mag \vs ${\langle V\rangle=19.24}$~mag and 
${\langle B-V\rangle=0.92}$~mag \vs ${\langle B-V\rangle=0.93}$ mag, again for 
Clementini \etal (2000) and OGLE, respectively. As can be seen the agreement 
between those two photometric data sets is also very good. Comparison of red clump 
stars is certainly much more sound because of two orders of magnitude 
larger number of stars. 

Summarizing, comparison of the OGLE {\it BVI} photometry with other 
observations indicates that in each band the agreement of the zero points is 
quite good. Because the OGLE calibrations were obtained on large number of 
photometric nights with use of wide range of standard fields they are 
certainly very reliable. We estimate the uncertainty of the zero points to be 
less than 0.02~mag. The OGLE-II maps constitute a huge set of secondary 
standards in the LMC line of sight. 

\Subsection{Completeness}
To estimate the completeness of detection of stars in the OGLE fields we performed 
similar set of tests as in the case of the SMC maps. For details the reader is 
referred to Udalski \etal (1998b). Table~4 presents results of these tests for 
three fields of different stellar density. As can be seen the completeness is high down to 
stars as faint as ${I\approx19.5}$~mag, ${V\approx20}$~mag and 
${B\approx20}$~mag. For fainter stars it gradually drops. The completeness is 
also a function of stellar density of the field. 
\MakeTable{c@{\hspace{6pt}}c@{\hspace{6pt}}c@{\hspace{6pt}}c
@{\hspace{6pt}}c@{\hspace{6pt}}c@{\hspace{6pt}}c@{\hspace{6pt}}c
@{\hspace{6pt}}c@{\hspace{6pt}}c@{\hspace{6pt}}c@{\hspace{6pt}}c}
{12.5cm}{Completeness of the LMC maps}
{
\hline
Stars &&\multicolumn{2}{c}{Completeness}&&\multicolumn{3}{c}{Completeness}&&\multicolumn{3}{c}{Completeness}\\
per bin & $B$  &  SC$\_$6 &  SC$\_$12 & $V$  & SC$\_$6 &  SC$\_$12 &  SC$\_$26 & $I$  & SC$\_$6 &  SC$\_$12 &  SC$\_$26 \\
\hline
~2 & 13.9 &  99.5 & 100.0 & 14.3 &  99.5 & 100.0 & 100.0 & 13.8 &  99.5 & 100.0 & 100.0 \\ 
~5 & 14.4 & 100.0 &  99.8 & 14.8 &  99.8 &  99.8 & 100.0 & 14.3 &  99.6 &  99.8 &  99.8 \\ 
~7 & 14.9 &  99.9 & 100.0 & 15.3 & 100.0 & 100.0 &  99.9 & 14.8 &  99.7 & 100.0 &  99.8 \\ 
10 & 15.4 & 100.0 &  99.6 & 15.8 &  99.3 &  99.8 &  99.7 & 15.3 &  99.2 &  99.9 &  99.3 \\ 
12 & 15.9 &  99.5 &  99.6 & 16.3 &  99.3 &  99.6 &  99.4 & 15.8 &  99.2 &  99.4 &  99.7 \\ 
15 & 16.4 &  99.3 &  99.6 & 16.8 &  98.9 &  99.7 &  99.6 & 16.3 &  98.7 &  99.5 &  99.3 \\ 
17 & 16.9 &  98.9 &  99.8 & 17.3 &  97.9 &  99.5 &  98.7 & 16.8 &  97.6 &  98.8 &  99.4 \\ 
20 & 17.4 &  98.9 &  99.2 & 17.8 &  95.6 &  98.5 &  98.8 & 17.3 &  95.5 &  98.0 &  98.3 \\ 
22 & 17.9 &  96.9 &  99.1 & 18.3 &  92.6 &  97.9 &  97.9 & 17.8 &  92.0 &  96.9 &  97.9 \\ 
25 & 18.4 &  95.4 &  98.5 & 18.8 &  86.6 &  96.0 &  96.6 & 18.3 &  89.0 &  96.4 &  96.5 \\ 
27 & 18.9 &  93.1 &  96.9 & 19.3 &  80.0 &  94.2 &  93.8 & 18.8 &  84.6 &  94.5 &  94.4 \\ 
30 & 19.4 &  83.7 &  93.0 & 19.8 &  68.7 &  90.3 &  89.3 & 19.3 &  74.2 &  90.8 &  85.9 \\ 
32 & 19.9 &  69.6 &  88.1 & 20.3 &  52.0 &  83.8 &  80.1 & 19.8 &  55.0 &  81.2 &  45.9 \\ 
35 & 20.4 &  47.8 &  78.3 & 20.8 &  28.5 &  71.7 &  59.1 & 20.3 &  23.0 &  51.3 &  10.3 \\ 
\hline
}

\Section{Color-Magnitude Diagrams}
Figs.~7--19 show {\it I} \vs $(V-I)$ color-magnitude diagrams (CMDs) of all 
LMC fields. They are presented to illustrate quality of data and potential 
usefulness of the OGLE LMC maps for studying many important astrophysical 
problems. About 25000 stars (\ie about 5--50\% of the total number) from each 
field are plotted in these figures for clarity. 

The CMDs presented in Figs.~7--19 reveal in great detail the most 
characteristic features of the LMC population of stars: main sequence stars, 
red giant branch stars, prominent red clump etc. The shape of the red clump 
can be an indicator of regions with non-uniform extinction in the LMC. For 
instance, red clump is severely elongated in the direction of reddening in the 
fields LMC$\_$SC16--18, while relatively round in the center of the bar 
(fields LMC$\_$SC2--6). 

\Section{Data Availability}
The {\it BVI} maps of the LMC are available to the astronomical community in 
the electronic form from the OGLE archive: 
\begin{center}
{\it http://www.astrouw.edu.pl/\~{}ogle} \\
{\it ftp://sirius.astrouw.edu.pl/ogle/ogle2/maps/lmc/}\\
\end{center}
or its US mirror
\begin{center}
{\it http://bulge.princeton.edu/\~{}ogle}\\
{\it ftp://bulge.princeton.edu/ogle/ogle2/maps/lmc/}\\
\end{center}

Also {\it I}-band FITS template images of the OGLE-II fields are included. 
Total volume of the compressed data is equal to about 0.7~GB. Usage of the data is allowed 
under the condition of proper acknowledgment to the OGLE project. 

We provide these data in the most original form to avoid any additional 
biases. For instance we do not mask bright stars which often produce many 
artifacts, but such a masking could potentially remove some interesting 
information on objects located close to bright stars. We do not remove objects 
which are located in overlapping areas between the neighboring fields. 
Cross-identification of these objects can be easily done based on provided 
equatorial coordinates. 

\Acknow{We would like to thank Prof.\ Bohdan Paczy\'nski for many discussions 
and help at all stages of the OGLE project. We thank Dr.\ D.S.~Graff for 
information about the systematic error in the early OGLE calibrations. We also 
thank Mr.~A.~Olech for carrying out part of the observations of the LMC in the 
1997 observing season. The paper was partly supported by the Polish KBN grant 
2P03D00814 to A.\ Udalski and 2P03D00916 to M.\ Szyma{\'n}ski. Partial support 
for the OGLE project was provided with the NSF grant AST-9820314 to 
B.~Paczy\'nski. We acknowledge usage of The Digitized Sky Survey which was 
produced at the Space Telescope Science Institute based on photographic data 
obtained using The UK Schmidt Telescope, operated by the Royal Observatory 
Edinburgh.}

\newpage
\begin{figure}[htb]
\FigCap{OGLE-II fields in the LMC. North is up and East to the left in
the Digitized Sky Survey image of the LMC.}
\vskip5mm
\FigCap{Differences of magnitudes of the same objects in the overlapping regions 
of fields LMC$\_$SC3 and LMC$\_$SC4.}
\vskip5mm
\FigCap{Differences of magnitudes of the same objects in the overlapping regions
of fields LMC$\_$SC14 and LMC$\_$SC15.}
\vskip5mm
\FigCap{Differences of magnitudes of the same objects in the overlapping regions
of fields LMC$\_$SC20 and LMC$\_$SC19.}
\vskip5mm
\FigCap{Standard deviation of magnitudes of objects in the field LMC$\_$SC6.} 
\vskip5mm
\FigCap{Standard deviation of magnitudes of objects in the field LMC$\_$SC12.}
\vskip5mm
\FigCap{Color-magnitude diagrams of the fields LMC$\_$SC1 and LMC$\_$SC2.}
\vskip5mm
\FigCap{Color-magnitude diagrams of the fields LMC$\_$SC3 and LMC$\_$SC4.}
\vskip5mm
\FigCap{Color-magnitude diagrams of the fields LMC$\_$SC5 and LMC$\_$SC6.}
\vskip5mm
\FigCap{Color-magnitude diagrams of the fields LMC$\_$SC7 and LMC$\_$SC8.}
\vskip5mm
\FigCap{Color-magnitude diagrams of the fields LMC$\_$SC9 and LMC$\_$SC10.}
\vskip5mm
\FigCap{Color-magnitude diagrams of the fields LMC$\_$SC11 and LMC$\_$SC12.}
\vskip5mm
\FigCap{Color-magnitude diagrams of the fields LMC$\_$SC13 and LMC$\_$SC14.}
\vskip5mm
\FigCap{Color-magnitude diagrams of the fields LMC$\_$SC15 and LMC$\_$SC16.}
\vskip5mm
\FigCap{Color-magnitude diagrams of the fields LMC$\_$SC17 and LMC$\_$SC18.}
\vskip5mm
\FigCap{Color-magnitude diagrams of the fields LMC$\_$SC19 and LMC$\_$SC20.}
\vskip5mm
\FigCap{Color-magnitude diagrams of the fields LMC$\_$SC21 and LMC$\_$SC22.}
\vskip5mm
\FigCap{Color-magnitude diagrams of the fields LMC$\_$SC23 and LMC$\_$SC24.}
\vskip5mm
\FigCap{Color-magnitude diagrams of the fields LMC$\_$SC25 and LMC$\_$SC26.}
\end{figure}
\end{document}